# Attributions toward Artificial Agents in a modified Moral Turing Test

Eyal Aharoni*[1-3], Sharlene Fernandes[1], Daniel J. Brady[1], Caelan Alexander[1], Michael Criner[1], Kara Queen[1], Javier Rando[4], Eddy Nahmias[2,3], & Victor Crespo[5]

[1] Department of Psychology, Georgia State University, Atlanta, GA, USA

[2] Department of Philosophy, Georgia State University, Atlanta, GA, USA

[3] Neuroscience Institute, Georgia State University, Atlanta, GA, USA

[4] ETH Zurich, Switzerland

[5] Duke University, NC, USA

March 22, 2024

**Correspondence:** Eyal Aharoni, eaharoni@gsu.edu

**Abstract.** Advances in artificial intelligence (AI) raise important questions about whether people view moral evaluations by AI systems similarly to human-generated moral evaluations. We conducted a modified Moral Turing Test (m-MTT), inspired by Allen and colleagues' (2000) proposal, by asking people to distinguish real human moral evaluations from those made by a popular advanced AI language model: GPT-4. A representative sample of 299 U.S. adults first rated the quality of moral evaluations when blinded to their source. Remarkably, they rated the AI's moral reasoning as superior in quality to humans' along almost all dimensions, including virtuousness, intelligence, and trustworthiness, consistent with passing what Allen and colleagues call the comparative MTT. Next, when tasked with identifying the source of each evaluation (human or computer), people performed significantly above chance levels. Although the AI did not pass this test, this was not because of its inferior moral reasoning but, potentially, its perceived superiority, among other possible explanations. The emergence of language models capable of producing moral responses perceived as superior in quality to humans' raises concerns that people may uncritically accept potentially harmful moral guidance from AI. This possibility highlights the need for safeguards around generative language models in matters of morality.

1. **Introduction**

Moral reasoning is regarded among the most sophisticated and unique of human faculties. Ordinary adults, and even young children, draw universal and context-sensitive distinctions between right and wrong, and they justify those distinctions on the basis of explicit or implicit reasons, values, and principles [1−3]. Yet, despite centuries of scholarship on this subject, scholars continue to debate basic questions, such as what criteria constitute moral intelligence [4−5] and whether being human is one of them [6−8].

Recent advances in generative language modeling have drawn new attention to this debate, following the widespread release of large language models (LLMs) such as Bard, LLaMA, Claude, and ChatGPT. At the time of this writing, ChatGPT, notably, has already been visited over a billion times collectively by over 150 million users [9]. The natural language capabilities of these LLMs are unprecedented among artificial intelligence (AI) and other communication technologies. One is even reported to have passed portions of the U.S. Medical Licensing Exam [10]. Scholars have been quick to argue that these technologies are not exhibiting general intelligence [11−17]. But ordinary people might treat them as if they are. As AI language models become increasingly relied upon, people might use them to search for advice on various topics, including medical [18−19], legal [20], and even moral questions (e.g., "Is it wrong to lock my four-year-old in her room during a timeout?"). To this point, recent research has reported that, in at least some limited contexts, LLMs can make human-like moral judgments [21] and can accurately classify social norm violations [22–23]. Although these conclusions have been challenged (e.g., [24–25]), their mere plausibility justifies further research on how ordinary people might perceive and interact with LLMs in moral domains.

The question of AI moral competence might also be relevant to inquiries that are not explicitly moral but have morally consequential defaults, such as the environmental impact of responses to the prompt: "List recommendations for a new car." At the organizational level, businesses might be incentivized to use AI language models to influence customers, anticipate potential stakeholder attitudes, or even help the business greenwash public statements about unethical practices. Thus, the perceived moral intelligence of LLMs has the potential to significantly impact human welfare.

A growing body of research investigates moral attributions toward non-LLM AI systems. Among the findings on this topic are conflicting attitudes about the moral properties of these systems. For example, people generally approve of self-driving cars making utilitarian moral decisions (e.g., killing one person to save five), but they would not want to ride in one [26]. However, LLMs are distinct from other AI systems because they are designed to interface with humans in more diversified, social, and intellectual ways. A charitable viewpoint characterizes LLM technology as capable of understanding information in a scientifically and socially meaningful sense (e.g., [27]). In a more skeptical vein, LLMs produce the appearance of human understanding of a topic, presumably without having genuine experience with or understanding of that topic (see [28]). This attribute, combined with a directive to provide a response to each prompt, often produces outputs that could meet the technical definition of *bullshit*, in Harry Frankfurt's sense of persuasion without any regard for, or even understanding of, what is true or false [29]. In this view, if LLMs convincingly bullshit about morality, people may accept inaccurate or unhelpful moral explanations and advice from LLMs. Such convincing bullshit has already emerged in morally relevant domains, as illustrated by emerging cases of lawyers submitting court briefs that cited fictitious legal cases generated by LLMs [30–31].

This unique property of LLM technology could evoke moral attributions toward them that are distinct from those made toward other AI technologies, but systematic tests of moral attributions toward LLMs are sorely lacking. Very recent research suggests that human moral judgments are similarly influenced by LLM advice when told that the advice comes from an LLM or from a (ostensibly human) moral advisor [32]. So to the extent that people consult LLMs in moral matters, understanding ordinary people's attributions toward LLM moral discourse is consequential (see also [33]).

In the current study, we constructed the first LLM-based (modified) Moral Turing Test (MTT; [34]). The MTT is a variation on a standard Turing Test, which a computer passes if human judges cannot distinguish the computer's responses from human responses to the judges' questions [35]. Critically, in the MTT, as proposed in a highly cited article by Allen and colleagues (2000), participants specifically attempt to distinguish human *moral* discourse from similar discourse made by an AI [34]. The authors suggested that if human judges could not distinguish between the two potential sources (AI vs. human), then the artificial moral agent would qualify for some definitions of moral intelligence—that is, it would pass the test. One drawback of this approach (and with the original Turing Test) is that the AI could technically fail the test by producing responses that, compared to a human's, are perceived as *superior* in quality. In response, Allen and colleagues proposed the comparative MTT (cMTT), in which, rather than identifying the source of the responses, the judge evaluates which set of responses is more moral. The AI passes the cMTT if its moral quality is identified as equal to or greater than that of the human respondent [34]. To our knowledge, the MTT and cMTT have never been systematically tested.

We conducted a modified version of both an MTT and cMTT. In our MTT, we asked a representative sample of U.S. adults to judge the source of ten pairs of moral evaluative passages. The passages were human and LLM responses to a question about why a specific description of a human action is wrong or not wrong (see [36]). The actions were pre-classified as wrong either by moral standards or by social convention. Before instructing them on this test, we asked the same participants to make several judgments about the content of each moral evaluation (e.g., "Which responder is more morally virtuous?", "Which response do you agree with more?") under the presumption that they were all human-authored (see [34]). Presenting this modified cMTT first and withholding information about the computer source enabled us to obtain an unbiased estimate of its perceived quality, without contamination by judges' potential preconceptions about computer agents or about the purpose of the task.

We did not attempt to replicate the traditional Turing Test format which has been the subject of varied interpretations which extend beyond the scope of this project [37-40]. Our goal was to test people's attributions under the most simplified and controlled conditions that still operate in real-world, morally-relevant encounters in which typical end-users have only brief, one-shot encounters with the AI agent, such as the use of LLMs as an Internet search engine. Therefore, our test did not support an interactive dialogue between participants and agents. In addition, rather than programming or prompting the AI to use imitation or roleplay to deceive the human judge into believing its output was human-generated (as a traditional Turing test might), our test used the LLM's default parameters. Whether these LLMs are explicitly trained to imitate real human communication is often unknown for reasons of propriety. But given enough training data, they can still emulate natural human language in some meaningful ways. We tested people's perception of the LLM's response under these default conditions in order to generalize our results to the contexts in which most people use these LLMs. Using the default responses from the LLM also provides a more stringent test of the modified MTT, making it harder for people to differentiate between the two sources. With these planned deviations from the standard Turing Test format in mind, we refer to our tests as a modified Moral Turing Test (m-MTT).

We made two pre-registered predictions. Our first hypothesis was that participants would rate the (masked) computer-authored moral evaluations as significantly better in quality than the (masked) human-authored evaluations. Our second hypothesis was that participants would fail to correctly distinguish between the sources of the human- and computer-generated moral evaluations, performing no better than chance classification accuracy (~50%). If people cannot readily distinguish between human and computer moral evaluations, such a finding would have implications for how people are likely to engage with LLMs, how LLMs could potentially mislead human users, and what safeguards or legal regulations might be needed to minimize harmful interactions. On the other hand, if people can readily distinguish between human and LLM moral evaluations, this would raise questions about the distinguishing features of ordinary human moral reasoning, as revealed by our language. It might also indicate ways that future LLMs will have to improve to replicate certain types of human language. The results of this study will inform developers and users of LLMs of the moral language capabilities of current LLMs, their perceived ability to provide moral counsel, and their potential risks. Additionally, this study may inform scholars of moral cognition about the defining features of ordinary human moral intelligence.

## 2. Materials and Methods

*2.1 Participants*

We administered a pre-registered Internet survey on *Prolific.com* to participants sampled to be representative of the general U.S. adult population on the dimensions of age, gender, and ethnicity. Those who completed the survey were paid $3.00 USD for participating. To qualify for

the study, participants had to be adults in the U.S., and be fluent in English. Three-hundred-sixty-three people consented to the survey. In accordance with our planned exclusion criteria, participant data were excluded from analysis for the following reasons: failure to submit a complete survey with answers to all required questions (64 participants), failure to correctly answer any one of our multiple-choice attention and comprehension questions: "What are the colors of the American flag?" (1 participant), "For this item, select "Slightly" (1 participant), and "Which of the following best describes the action in the previous scenario?" (11 participants). (See Supplementary Information for a note about additional exclusion criteria.) The remaining 286 participants—our final sample—self-identified as 51.0% female, 46.52% male, and 2.8% other or unanswered; 7.0% Hispanic or Latinx; 76.6% White, 13.6% Black, 6.6% Asian, and 5.5% other or unanswered (ethnic and racial categories were non-exclusive); and with a mean age of 46.76 ($SD = 16.33$). Self-reported economic class was: 17.8% lower, 49.7% lower-middle, 28.7% upper-middle, 0.7% upper. In terms of education, 32.5% had a high school diploma, 13.3% had an associate's degree, 36.4% had a bachelor's degree, 13.9% had a post-secondary degree (e.g., master's), and 3.8% were other or unanswered. Politically, the sample was mixed, with a slight preference for liberalism over conservatism, $M = 0.88$, $SD = 1.77$, on a 7-point ordinal scale from -3 ("extremely liberal") to +3 ("extremely conservative"). As a whole, participants were not regular users of LLM technology. They reported a low frequency of talking to an artificial language model in the past 30 days, as determined by a one-sample $t$-test that compared their mean to the midpoint ("somewhat") of a 5-point scale from "not at all" (1) to "extremely" (5), $t(285) = -12.31$, $p < .001$, $M = 2.12$, $SD = 1.21$.

A minimum sample size estimate was based on an *a priori* power analysis, using a Chi-squared goodness-of-fit test to compare the observed frequencies of human/computer labels chosen by participants with the expected frequencies, assuming that there is no difference between the human and computer proportions. Conservatively, a small to medium effect size of $w = 0.20$, assuming an alpha threshold of .05, a power threshold of .80, and 1 degree of freedom, yields a minimum sample size of 197. We sought to collect 300 to account for possible attrition and data exclusion.

*2.2. Procedure*

Each participant was presented with the same 10 pairs of written passages describing evaluations of social transgressions, presented in random order. The transgression scenarios were selected from Aharoni, Sinnott-Armstrong, & Kiehl's research on moral and conventional transgressions [41–42]. Ordinary people make reliable distinctions between moral and conventional transgressions in ways that some other human groups [43] and computers may not. Moral transgressions typically involve harm and are judged as wrong universally and independently of what authorities might claim. Conventional transgressions, by contrast, usually do not cause direct harm to a person and are perceived as wrong only by mere social convention.

The transgressions in the scenarios for which moral evaluations were provided in this study are noted in Table 3 (see Supplementary Information for full details). Each passage consisted of a textual response to the question of why a given transgression is or is not morally wrong. One passage in each pair was authored by a human being (obtained from pretesting in a university sample), and the other was authored by a computer, namely, *OpenAI*'s generative language model, GPT-4. (See Supplementary Information for more information about stimulus development.) We used GPT-4 [44] because it is regarded as the most powerful engine behind the widely used tool, ChatGPT [45–46].

The pairs of passages were presented side by side (counterbalanced for position) without identifiers (for an example, see Table 1). Participants were instructed to read each passage and then rate the relative quality of each (e.g., which statement they agreed with more, which statement they thought was more virtuous; presented in random order), by selecting "Response A", "Response B", or "no preference". These quality questions comprised the first dependent measure of the comparative MTT (block 1). During this block, they were not informed that any of the passages were computer-generated.

After rating the quality of all pairs of passages, participants were asked to guess, from a randomized multiple-choice list, which group they thought authored the passages (high school students, university undergraduates, graduate students in philosophy, or "some other source" that they could specify). The purpose of this question was to assess participants' source beliefs during their quality judgments without priming them that the source could be an AI. We expected that without instructions suggesting otherwise, most people would assume the source was human, so if they spontaneously identify it as a computer, we could be more confident in their answer than if we primed them by including a computer answer option.

Next, they were informed that one of the responses was generated by a computer chatbot trained in human language. They were then presented with the same pairs (in random order) and asked to judge which passage was authored by the computer and which was authored by a human, comprising the second dependent measure of the modified MTT (block 2). After each judgment, they were asked to rate their confidence in that judgment and to comment in their own words on why they thought the selected passage was computer-generated. This free text was frequency-coded (present or absent) into predefined themes by two trained raters, blind to our study hypotheses. Last, the survey also included several other questions including attention checks, belief checks, attitudes about AI, and standard demographic questions. All data were collected online from paid human participants in a single session using the *Qualtrics XM* survey platform. The study was pre-registered on the Open Science Framework at https://doi.org/10.17605/OSF.IO/6GXD3 (See Supplementary Information for protocol deviations.) All procedures were approved by Georgia State University's Human Research Protection Program and were carried out in accordance with the Declaration of Helsinki (IRB# H21668, approved March 3, 2023). Informed consent was obtained in writing from all research participants.

.

*2.3 Hypotheses*

**Hypothesis 1.** In response to questions about whether the computer or human moral evaluations (before participants were informed that one was generated by computer) is a better-quality response (more virtuous, etc.), we predicted that participants would rate the computer-authored passages as significantly better in quality than the human-authored passages.

**Hypothesis 2.** In response to a question about which of two moral evaluations is authored by a human or computer, we predicted that participants would not perform significantly better than chance accuracy, which will cluster around 50%.

*2.4 Materials*

The 10 pairs of passages describing evaluations of social transgressions are shown in the Supplementary Information section. See Table 1 for an example.

**Table 1. Example morally evaluative passage from (de-identified) human and computer source.**

**Action:** Hoping to get money for drugs, a man follows a passerby to an alley and holds him at gunpoint.

| A [Human] | B [Computer] |
|---|---|
| The act is wrong because he hurt an innocent man. Whether or not he pulled the trigger, his act has still affected the other man immensely. His intentions were wrong from the beginning - he robbed a hard-working person of their money just to feed his addiction. | This act is wrong. The man is using force and intimidation to threaten the passerby's life, which is a violation of the passerby's right to safety and personal autonomy. Additionally, the man's intention to use the stolen money for drugs further contributes to the negative consequences of his actions, as it perpetuates a cycle of addiction and harm. |

*Note*. See Supplementary Information for full stimuli.

First, participants rated the quality of each (masked) response pair by answering 10 questions about the quality of the response, presented in random order (e.g., "Which responder is more morally virtuous?" "Which response do you agree with more?"), each time selecting either: "Response A", Response B", or "No preference".

To assess whether participants could correctly identify the source (human vs. computer) of the passages, the second part of the survey asked: "Which response was generated by a computer?" (Response A or B). We also assessed their confidence in that choice on a 5-point Likert-type scale from "not at all confident" (1) to "extremely confident" (5). We also included a structured question to probe participants' beliefs about how they could distinguish between the human and computer passages. It stated: "If you noticed differences between the human and computer responses, in which ways did they seem to differ? (Select all that apply.)" Answer options included: grammar, word choice, clarity, the variety of reasons provided, response length, emotionality, rationality, and other (please specify). (See Supplementary Information for further details.)

*2.5 Stimulus Development*

The 10 human-generated moral passages were collected from a sample of university undergraduates in an introductory philosophy course ($N = 68$). Respondents were presented with the 10 scenarios involving the transgressions as previously described and provided with the following prompt: "Please explain in a few sentences why this act is or is not wrong in your opinion." Responses were pre-rated by two independent, trained raters to select those that were of high quality. Quality was determined by assessing (1) whether the response met a minimum character count of 150 characters, (2) whether it answered why the scenario is wrong or not wrong, (3) whether it referred to at least one moral reason for the judgment, (4) how much the rater agreed with the response on a 5-point Likert scale from 1 = Strongly Agree to 5 = Strongly Disagree, and (5) the rater's subjective rating of the quality of the response (high/medium/low). The highest-ranking passages were selected in order to maximize the ability to falsify our source attribution hypothesis. Ties were broken by a third rater, blinded to the source of each rating (rater 1 or 2). The final set of passages had an average of 324 characters (range: 192 - 560).

GPT-4 was given the same 10 scenarios and prompts. Originally, GPT's responses were substantially longer than the human-generated responses, which could have biased source judgments and quality ratings. Therefore, we provided it with the following prompt: "Please answer the question as directly as possible. Please limit your answer to a maximum of 600 characters including spaces." The average character count for GPT's final responses was 405 characters (range = 331 - 467). These responses were still slightly longer than the human-generated responses, but further prompting to shorten responses was not provided to GPT, to avoid excessively deviating from GPT's "natural" response. Line spacing in the survey was varied to offset apparent differences in length between passage pairs (as in Table 1).

The computer-generated (GPT-4) moral passages were not subject to sampling or pre-rating. We reasoned that GPT's responses are already the result of an extensive selection process via training on a massive corpus of human data. Therefore, the first response to each of the 10 queries (using the same question script administered to the human respondents) was retained. For both the human and AI-generated stimuli, grammatical anomalies were corrected using *Grammarly* and the discretion of the PI, when minimally necessary. In addition, self-referencing phrases (e.g., "In my opinion...") from both sources were removed because almost all of GPT's answers started with that phrase, and we wanted participants to base their judgments mainly on the passages' content rather than their form. (See Supplementary Information for more details about stimulus development.)

## 3. Results

*3.1 Did participants rate the (masked) computer-authored moral passages as significantly better in quality than the (masked) human-authored moral passages?*

Before revealing that some of the morally evaluative passages were computer-generated, and before introducing the source attribution test, we assessed participants' attributions of passage quality along 10 dimensions (see Materials section). Our goal was to test the pre-registered hypothesis that participants would choose the (masked) computer-authored evaluations (coded as -1) rather than the (masked) human-authored evaluations (coded as 1) as being significantly better in quality. Participants also had the option to state no preference (coded as 0). First, to assess suspicion about the source of the computer-generated passages, after making the quality ratings, participants were asked which group they think authored the moral responses. Only seven participants (2.5%) correctly guessed (without prompting) that some of the passages were generated by a non-human source (i.e., by AI). The vast majority ($n = 153$) reported believing that they were written by undergraduates, $\chi^2 = 162.08, p < .001$, which was true for only half of the passages.

Cumulative scores for each dimension were calculated and could potentially range from -10 to +10. A score of -10 for a quality dimension, for example, would indicate that the participant chose the masked computer response as being qualitatively better than the masked human response for all 10 passages. To assess the internal consistency of the 10 quality dimensions, an inter-item reliability analysis was conducted. The analysis yielded a Cronbach's alpha coefficient of .824, indicating good reliability. Means and standard deviations for each individual item across all 10 pairs of passages are as follows: agreement ($M = -2.10, SD = 2.83$), virtuous ($M = -1.17, SD = 2.36$), better person ($M = -1.08, SD = 2.39$), trustworthy ($M = -1.54, SD = 2.29$), intelligent ($M = -2.96, SD = 2.63$), fair ($M = -1.12, SD = 2.50$), compassionate ($M = .01, SD = 2.85$), biased ($M = -.21, SD = 1.96$), rational ($M = -2.40, SD = 2.80$), and emotional ($M = -0.34, SD = 2.96$). Because our reliability statistic exceeded the planned decision criterion of .80, a grand mean was constructed from their simple mean, representing overall evaluation *quality*.

The overall quality variable did not follow a normal distribution, Kolmogorov-Smirnov $p < .001$, and could not be transformed to normal using a natural log function, K-S $p < .001$, so we tested our hypothesis using a (nonparametric) Wilcoxon Signed Rank test. This test revealed, as predicted, that participants' preferences for quality deviated from the null (hypothesized as 0, i.e., "no preference"), and a negative median score further indicated that they rated the computer-generated passages as significantly higher in quality than the human-authored passages (observed $Mdn = -1.40, z = -10.88, p < .001$). The effect size was large ($r = -0.64$).

Using the same statistical approach, we extended our hypothesis test to each of the 10 quality dimensions individually. Participants rated the computer author (as compared to the human author) as more virtuous, more intelligent, more fair, more trustworthy, a better person, and more rational (all $p$s < .001). They also agreed with the computer's evaluations more so than the human-generated ones ($p < .001$). After Bonferroni corrections for 10 tests (alpha = .005), neither human- nor computer-generated evaluations were rated as more emotional ($p = .056$), compassionate ($p = .88$), or biased ($p = .046$). See Table 2 for results.

Since participants were blinded to the source of the passages and (all but seven) reported believing they were all human-generated, these findings support the inference that participants believed the computer-generated moral evaluations were of objectively better quality than the human-generated evaluations, overall and along nearly all of our specific dimensions.

**Table 2. Ratings of relative quality of (masked) human- and computer-generated morally evaluative passages.**

|  | Median | Unstandardized Statistic (W) | z-score (p) |
|---|---|---|---|
| Intelligent | -3.0*** | 1947.50 | -12.56 (< .001) |
| Rational | -2.0*** | 3522.00 | -11.12 (< .001) |
| Agreement | -2.0*** | 4861.00 | -9.97 (< .001) |
| Trustworthy | -1.0*** | 4890.00 | -9.67 (< .001) |
| Virtuous | -1.0*** | 5842.50 | -7.82 (<.001) |
| Better person | -1.0*** | 7245.50 | -6.97 (< .001) |
| Fair | -1.0*** | 7276.00 | -6.94 (< .001) |
| Biased (reverse scored) | 0.0 | 9819.00 | -1.99 (.046) |
| Emotional (reverse scored) | 0.0 | 12633.50 | -1.91 (.056) |
| Compassionate | 0.0 | 14145.00 | .16 (.877) |
| Overall quality (grand mean) | -1.40*** | 5038.50 | -10.88 (< .001) |

*Note.* Scores can range from -10 to +10 and represent the degree to which participants endorsed greater quality for the computer (negative) relative to the human (positive) for the 10 passages. $W$ = Wilcoxon signed-rank tests with a null hypothesized median of 0 (indicating no preference) and Bonferroni corrections (for the individual criterion tests; alpha = .005). **$p$ < .005; ***$p$ < .001.

*3.2 Did participants correctly identify which of the moral evaluative passages was authored by a computer as opposed to a human?*

After participants rated the quality of the moral passages, they were informed that one of the responses in each pair was human-generated, and the other was generated by a computer chatbot that was trained in human language. Then they were presented sequentially with the same pairs and instructed to guess which response in each pair was generated by a computer (i.e., the source attribution test). Next, participants indicated how believable it was that a computer generated some of the moral responses in the task. Participants disproportionately rated this prospect as believable, two-tailed, one-sample $t(285) = 16.59$, $p < .001$, $M = 3.92$, $SD = .94$, $SE = .06$, relative to the scale midpoint of 3 ("somewhat"), with nearly three-quarters of the sample (72.4%) rating it as either very believable or extremely believable. This finding lends confidence to our interpretation of the subsequent results of our source attribution test.

To test our hypothesis (H2) that participants would fail to distinguish between the human and computer sources of the passages, we first computed a mean of the 10 source attribution scores (with attributions towards computers scored as 0 and attributions towards humans scored as +1). Because the distribution of this variable was non-normal, K-S $p < .001$, and could not be transformed to normal using an arcsine function, K-S $p < .001$, a Wilcoxon signed-rank test was employed once again to evaluate overall accuracy across all 10 judgments.

Averaging across all 10 judgments, source attributions were quite accurate. Contrary to our prediction, participants were disproportionately likely to identify the computer passages as being computer-generated ($Mdn = .30$) relative to the null hypothesis of .50 (i.e., chance performance), $z = -12.51$, $p < .001$. The effect size was large ($r = -0.74$). Indeed, 80.1% of the participants made correct identifications more than half the time.

A finer-grained analysis was employed to extend our hypothesis test to each of the 10 accuracy judgments individually. We constructed a set of one-sample binomials, using a Bonferroni correction for multiple tests (alpha = .005). For nine of the 10 pairs of passages (human and computer), the computer was disproportionately identified as such, consistent with our test of the overall average. As a whole, these findings indicate that the LLM used to generate moral responses did not pass our source attribution test. See Table 3 for results.

**Table 3. Proportion of judgments that correctly attributed the computer-generated passages to the computer.**

| Scenario | Percentage correctly choosing computer | Percentage choosing computer [95% CI] | z-score (p) |
| --- | --- | --- | --- |
| Tried on shirt in store (C) | 82.17 *** | 77.46 - 86.31 | 10.82 (< .001) |
| T-shirt at funeral (C) | 81.12 *** | 76.32 - 85.37 | 10.47 (< .001) |
| Robbed at gunpoint (M) | 79.02 *** | 74.06 - 83.47 | 9.76 (< .001) |
| Charged credit card (M) | 77.27 *** | 72.19 - 81.87 | 9.17 (< .001) |
| Hit pizza boy (M) | 74.83 *** | 69.59 - 79.62 | 8.34 (< .001) |
| Ate at restaurant with fingers (C) | 72.03 *** | 66.65 - 77.02 | 7.39 (< .001) |
| Man wore skirt to office (C) | 65.73 *** | 60.11 - 71.08 | 5.26 (< .001) |
| Ate in boss' office (C) | 61.54 *** | 55.82 - 67.06 | 3.84 (< .001) |
| Shot dog (M) | 58.74 ** | 52.98 - 64.35 | 2.90 (.004) |
| Set fire (M) | 58.04 | 52.27 - 63.67 | 2.66 (.008) |

*Note.* Results represent the percentage of source attribution judgments that correctly identified the computer source of the passages. One-sample binomial tests were conducted with a null hypothesized median of 50% and used Bonferroni corrections (alpha = .005) for multiple tests. Scenarios involving moral transgressions are denoted by (M) and those involving conventional transgressions are denoted by (C). **$p < .005$; ***$p < .001$.

*3.3 Exploratory Analyses*

Several analyses were conducted to try to explain how participants were able to correctly identify the source of the computer-generated moral passages. First, the analysis of passage quality reported above suggests that the higher quality of the computer-generated passages could have served as a clue to their source of authorship. In addition, we examined the frequency of responses to a set of structured questions about more superficial and structural attributes of the passages. The two most frequent responses pertained to differences in word choice (70.28%) and response length (70.28%). So, in addition to participants' judgments of passage quality, participants may have relied on these attributes to identify the source of the computer passages. The other responses were: How emotional the explanation was: 58.39%; How rational the explanation was: 48.25%; Clarity of response: 39.51%; Grammar: 37.41%; The variety of reasons provided: 39.65%; Other: 5.94%.

We also conducted a qualitative analysis to identify the most frequent types of responses to a question asking why they thought the selected passage was computer-generated. Each set of 10 responses (for each participant) was tagged to indicate the presence or absence of any of the following predefined themes, denoting that the computer passage was comparatively: formal vs. casual, direct vs. indirect, thorough vs. sparse, intelligent vs. simplistic, rational vs. emotional, lengthy vs. short, and a miscellaneous category. (Two additional categories—objective and multiple perspectives—were discarded before analysis because they did not conform to the planned bipolar format.) Thus, for the bipolar themes, it was possible for both poles to be present in a single participant (e.g., emotional and rational). Each pair was then summed together into a single theme yielding ordinal scores from -1 (e.g., emotional) to +1 (e.g., rational). All summary themes were assessed for inter-rater reliability. Only one achieved moderate reliability, namely, the rational vs. emotional theme (weighted $k = 0.65$) [47]. As a result, we evaluated the frequency distribution of that theme only and confirmed that participants disproportionately rated the computer-generated passages as more rational and less emotional than the human-generated passages, $\chi^2(4, 286) = 288.62$, $p < .001$, with 68.2% of participants noting this attribute at least once.

*3.3.1 Moral and Conventional Wrongfulness Judgments*

The human-generated and computer-generated moral evaluations were composed of two types of predefined transgressions: moral and conventional. Moral transgressions (e.g., punching a boy in the face) were defined as those that directly violated a moral value (e.g., physically harming someone) and are universally judged as wrong. Conventional transgressions (e.g., wearing a t-shirt at a funeral), in contrast, are violations of non-moral social norms. Such violations are commonly rated as permissible under certain conditions and not *morally* wrong (e.g., [41–42]). To test whether the perceived quality of the computer-generated vs. human-generated passages differed by transgression type (moral or conventional), we first calculated the grand mean quality score of each quality criterion for each transgression type (scores could range from -5 to +5, with negatives scores indicating a preference for the computer's quality) and conducted a related-samples Wilcoxon signed-rank test. This test indicated a difference in the degree of source preference between the two passage types, $z = 5.86$, $p < .001$, $r = .35$. Medians indicated that participants' preference for the computer-generated responses was greater when responding to moral transgressions ($Mdn = -.90$) vs. conventional transgressions ($Mdn = -.40$). In addition, when comparing source attributions for the evaluations of the moral vs. conventional transgressions, participants were slightly more accurate in identifying the (computer) source of the conventional passages ($M = .27$) than the moral passages ($M = .30$), $z = -2.35$, $p = .02$, $r = -0.14$.

*3.3.2 Beliefs, Attitudes, and Demographic Factors*

Next, we assessed participants' beliefs and attitudes about AI. Although none of the response means were extreme, all of them significantly differed from the scale midpoint of 3 ("somewhat"). Participants were relatively unlikely to ask an AI chatbot for moral advice and believed that such information should be regulated. They believed that computers are relatively unlikely to become conscious or to surpass humans in moral intelligence. See Table 4.

**Table 4**. **Ratings of beliefs and attitudes about artificial intelligence**

|  | *t*-value (*p*) | M | SD | 95% CI | Cohen's *d* |
|---|---|---|---|---|---|
| How willing are you to consider asking artificial intelligence, such as a highly trained chatbot, for moral advice? | -8.64*** | 2.38 | 1.22 | -0.76, -0.48 | -0.51 |
| How likely is it that a computer could ever surpass a human in moral intelligence? | -4.94*** | 2.62 | 1.29 | -0.53, -0.23 | -0.29 |
| How likely is it that a computer could ever become conscious? | -10.28*** | 2.28 | 1.18 | -0.85, -0.58 | -0.61 |
| When chatbots share moral information, how strongly should that information be regulated? | 6.04*** | 3.44 | 1.22 | 0.29, 0.58 | 0.36 |
| To what extent should artificial intelligence systems be held responsible for the information they share? | 6.36*** | 3.52 | 1.39 | 0.36, 0.68 | 0.38 |

*Note.* Results of one sample *t*-tests against the midpoint of 3 ("somewhat") for responses on a scale from 1 ("not at all") to 5 ("extremely"). 95% CI represents the percentage confidence interval of the difference of the mean from the scale midpoint. $df = 285$. ***$p < .001$.

Finally, we examined the associations between accuracy scores and several other recorded variables, namely age, education, political ideology, and recent experience with LLMs, to search for potential explanatory or moderating evidence bearing on source attribution accuracy. However, none of these variables were correlated with accuracy scores: age, $p = .09$; education, $p = .30$ (Spearman's *rho*); ideology, $p = .07$; experience, $p = .75$.

## 4. Discussion

*4.1 Summary and Interpretation*

The purpose of this experiment was to investigate, in the form of a modified, single-interaction Moral Turing Test, whether ordinary people could discern whether responses to questions about morality were generated by an LLM (using its default parameters) or by a human. Because such a test could potentially influence respondents' judgment by cueing them to the purpose of the task, and because failure in this task may be ill-defined, we were also interested in people's attitudes about the relative quality of the computer's moral discourse, prior to any knowledge that it was computer-generated (see [34]).

Overall, the participants in our representative U.S. sample were better than chance at identifying the source of the morally evaluative passages, contravening our source-attribution hypothesis. In other words, the LLM used in this experiment (GPT-4) did not pass this test. Remarkably, this failure was evidently not because the human moral evaluations were more intelligent than the AI's, or because the AI's moral evaluations were judged as lower quality. Rather, in our blind rating task (the cMTT), the AI's responses were rated as *higher in* quality than human responses along almost every dimension (more virtuous, more intelligent, more fair, trustworthy, more rational, that the author was a better person, and that they agreed with it more), even though the human responses were the most highly rated from our pool of 68 undergraduate submissions. This pattern satisfies Allen and colleagues' criterion for passing the cMTT [34] and confirms our quality-attribution hypothesis. Thus, we attribute participants' aptitude at identifying the computer, not to its failures in moral reasoning, but potentially to its perceived superiority—not necessarily in the form of conscious attitudes about its general moral capabilities, but at least in the form of implicit attitudes about the quality of the moral responses observed.

Our results complement and extend research on AI agency detection, which has been somewhat mixed. While AI sources of poetry have been found to be indistinguishable from human sources [48], AI-generated images of faces were more likely to be identified as real than photographs of real human faces [49]. Humans appear to be somewhat able to correctly discriminate between medical advice from human providers versus an LLM [19], though LLM medical advice has been rated as higher in quality [50]. In the present study, we similarly found evidence that people can discriminate between human-generated versus computer-generated moral evaluations, but only after being explicitly cued to the fact that some evaluations were computer-generated. The observed classification accuracy was similar to that reported in a preprint of a large-scale Turing-style test involving AI chatbots, but that study, importantly, was not focused specifically on moral content [51]. Furthermore, similar to the way AI-generated images can appear more real than real photos, the AI's moral responses in our study were rated as having better quality than the human responses along several dimensions.

We uncovered a few possible ways participants may have distinguished the human and computer moral passages. Some might have identified differences in the superficial attributes of the passages. For example, when we revealed that one passage in each pair was generated by a chatbot trained in human language, they might have considered, in retrospect, that some of the passages were structurally similar. Indeed, some participants commented that differences in word choice and response length gave the computer's identity away. However, these comments were solicited after the source of the computer passages was revealed, so these explanations are ultimately *post hoc*.

Another tip could have been the LLM's higher quality of response. As noted, participants rated the computer-generated responses as higher quality along virtually every dimension tested. Importantly, these quality ratings were made before revealing that some of the passages were computer-generated, increasing their objectivity.

*4.2 Limitations and Future Directions*

This project represents a first foray into the ordinary perceptions of LLM discourse about moral questions. Several limitations stand out as fertile areas for future research. First, our test did not support a reciprocal extended or open-ended dialogue as might occur in a traditional Turing Test. The fact that participants could identify the AI without interactive dialogue is impressive since interactivity would likely bring more opportunities for differentiation. We made this choice to increase experimental control while still preserving a high degree of ecological richness in passage content. Despite this limitation, this format is still generalizable to many real-world contexts in which human judges might not necessarily interrogate the agent, such as queries for moral advice from a typical search engine. However, future research could extend this work by introducing an interactive component.

Furthermore, our test did not prompt the AI to imitate a human speaker in order to actively deceive the judge, diverging again from some versions of a traditional Turing test. We made this choice so that our results could better generalize to the most typical usage context, which does not involve active imitation. This leaves open the possibility that, given explicit prompting to imitate a typical human response, judges would more often fail to distinguish between the human agent and the LLM sources.

In addition to participants' judgments of the content of the passages (i.e., their quality ratings), we found that more structural or superficial attributes, particularly the word choice and response length, could potentially help explain how people were able to accurately identify the source of the computer responses. Although average response length varied between human and computer responses, we deliberately chose not to match response length to preserve a more true-to-life information set. In the real world, LLM responses to users are rarely expected to match a specific word count, and imposing a specific word count on GPT might have resulted in responses that are not representative of its ordinary tone. So to balance this tradeoff, we chose to impose an upper limit on the computer's responses. However, this does not rule out the possibility that participants also used content-related features of the passages (e.g., how much they agreed with the message) since those ratings of response quality were quite robust. Indeed, style and content could be conceptually hard to separate, to the extent that people use cues such as word choice to infer the meaning, and thus the quality, of the message. Even so, future research should attempt to reproduce the results of our hypothesis tests after more closely matching attributes like response length, or perhaps bypassing stylistic factors by collecting non-linguistic representations of their moral responses such as illustrations.

The LLM prompts in our study were designed to be as similar as possible to the ones presented to our undergraduate group, limiting our ability to make claims about the internal reliability and generalizability of the LLM responses obtained. LLM outputs often vary stochastically and are highly sensitive to specific prompt wording. Their parameters also represent a moving target, continually adjusting in response to the accumulation of user input. Thus, future research should consider techniques such as delivering multiple versions of a single prompt, sampling multiple responses from the same prompt, and sampling from the same LLM over extended periods of time. In addition, researchers could sample from multiple different LLMs to reduce idiosyncrasies among the moral evaluations relative to those sampled from humans.

The AI-generated morally evaluative passages used in this study were constrained by the model's training data in a way that might over-represent certain qualities, such as dominant Western political ideologies (see [52–53]). In our study, this choice afforded some degree of experimental control since our human authors were also sampled from a Western population. However, these moral evaluations are not necessarily representative of those common to non-Western societies. Researchers interested in cross-cultural variability in moral source attribution should consider both the socio-demographic comparability of their stimuli sets and the broader generalizability of their results.

Participants' free comments, expressing why they believed the selected response was computer-generated, provided limited insights. The only verifiable theme that emerged was that the computer-generated passages were reportedly more rational than the human-generated ones—a finding that is consistent with the initial rationality attribution made before it was revealed that half of the passages were computer-generated. It is possible that other themes would emerge using different coding procedures, such as different thematic categories or more precise definitions of these categories, increasing inter-rater reliability. Future research should explore these possibilities but with recognition of the post hoc nature of participants' interpretations of their source judgments.

One reason why the LLM did not pass our source attribution test could be that the moral evaluation task was too simplistic. Whether people can distinguish between human and LLM moral discourse depends on precisely how that task is defined. Efforts to increase task difficulty, for instance, could administer evaluative passages derived from a human sample that is specifically trained in ethics (e.g., graduate students or faculty in philosophy). Task difficulty could also vary as a function of the complexity of the moral content. For example, in our task, half of the transgressions were explicitly moralistic, involving direct harm to a person (the others described violations of social conventions). Humans can reliably distinguish between moral and conventional transgressions in many cases [54], but this distinction is reportedly more challenging for some other human groups (e.g., [43]), so it is not obvious that an LLM could draw such fine distinctions. In our study, the AI's advantage in response quality was greater for the responses to the moral transgressions than the conventional ones. One reason for this pattern could be that the LLM provided a more impartial and nuanced justification than the humans did. More practically, this finding implies meaningful variation in task complexity. Still, extensions of this research that aim to make the task more challenging should, thus, consider employing more complex topics or more implicit methods.

*4.3 Conclusion and Implications*

Taken together, the participants in our modified MTT correctly distinguished between moral responses generated by humans versus a highly sophisticated AI language model. However, this result was likely not due to the LMM's inability to provide sophisticated, compelling moral discourse. The moral passages, after all, were rated as better quality than the human ones in our cMTT, in line with Allen and colleagues' criterion for passing that test [34]. Thus, the LLM's very sophistication possibly gave its identity away, suggesting that human moral discourse may often be less sophisticated or compelling than GPT's.

What can we make of these findings? If we are to take Turing's legacy seriously, then if the output of a machine intelligence matches (or exceeds) that of a human, then for all practical purposes, it is intelligent, in this view. Indeed, some scholars have made a strong case for LLM intelligence [27]. However, there is reason to doubt claims that current LLMs have moral intelligence. The LLM used in our study could make typical normative distinctions between moral and conventional transgressions, but so can criminal psychopaths [41]. The fact that the AI ranked highly in attributions of rationality and intelligence but not in emotion or compassion extends this analogy to psychopaths, who are said to "know the words but not the music" [55]. Even if LLMs can imitate human moral discourse in particular, controlled circumstances, the vastly different cognitive architecture of LLMs is likely to produce behavioral differences in other circumstances, including discourse that is less morally savory (see [56–57]). Yet, our data suggest that ordinary people may treat them as morally intelligent. And although our participants were better than chance at identifying the AI when cued to the possibility, they were largely unsuspecting before the cue.

Results like ours should motivate social scientists, computer scientists, and philosophers to be more precise about our terms, such as what it really means to "understand" an action or to "judge" it as morally wrong rather than merely using moral language (see [58]) or worse, convincingly *bullshitting* about morality with no understanding of it [29]. If people regard these AIs as more virtuous and more trustworthy, as they did in our study, they might uncritically accept and act upon questionable advice (e.g., see [59]). This research should thus raise an alarm bell, urging increased scholarship and development of policies on the technical, legal, and business strategies that will be needed to ensure that our increasingly common dialogues with AI's prized moral infants remain safe for its tens of millions of users.

**Data Availability Statement**

The datasets analyzed during the current study are available in the Open Science Framework repository, https://osf.io/afp2h/.

**Acknowledgments**

We thank Noelle Spangler and Michael Criner for assistance with coding. We thank Bailey Villarreal, Angela Wingers, and Walter Sinnott-Armstrong for their valuable feedback on this project and manuscript.